%% file: Paper.tex
\def\nv{\vec{\chi}}
\def\kk{\vec{q}}
\def\nv{\vec{\chi}}
\def\vR{\vec{R}}
\def\etal{et al.}
\documentclass[final,times,3p]{elsarticle}
\usepackage[colorlinks]{hyperref}
\usepackage{graphics}
\usepackage{graphicx}
\usepackage{subfigure}
\usepackage{amssymb}
\usepackage{epsfig}
\usepackage{epsf}
\usepackage{color}
\input{macro_library}

\begin{document}

\begin{frontmatter}
\title{\Large\bf{Disorder induced lifetime effects in binary disordered systems : a first principles formalism and an application to doped Graphene}}
\author{\bf Banasree Sadhukhan$^\dagger$, Subhadeep Bandyopadhyay$^\ddagger$, Arabinda Nayak$^\dagger$ and Abhijit Mookerjee}
\address{$^\dagger$ Department of Physics, Presidency University, 86/1
College Street, Kolkata 700073, India\\  $^\ddagger$ Department of Solid 
 State Physics,
Indian Association for the Cultivation of Science,
2A-2B, Raja Subodh Kumar Mullick Road, Jadavpur.
Kolkata 700032, India}
\address{Professor Emeritus,
Department of Condensed Matter and Materials Science, S.N. Bose National Centre for Basic Sciences, JD-III, Salt Lake, Kolkata 7000098,India.}

\vspace{0.2cm}
\begin{abstract}
\noindent In this work the conducting properties of graphene lattice with a particular concentration of defect (5\% and 10\%) has been studied. The real space block recursion  method introduced by Haydock \etal\ has been used in presence of the random distribution of defects in graphene. This Green function based method is found more powerful than the usual reciprocal based methods which need artificial periodicity.  Different resonant states appear because of the presence of topological and local defects are studied within the framework of Green function.
\end{abstract}
\end{frontmatter}

\section{Introduction}

 In 1928 Felix Bloch \cite{bloch} introduced a theorem which
immediately started an avalanche of activity on the band structure of electrons in crystalline solids. It provided   a very useful technique for solving the electronic part of the differential equation (later formally derived by Kohn and Sham \cite{hk,ks}). Bloch's ideas were primarily to find electronic properties of crystalline solids  based on 
the earlier works of three mathematicians : George William Hill (1827) \cite{hill}, Gaston Floquet (1883) \cite{floq} and Alexander Lyapunov  (1928) \cite{lyap}.  Bloch  \cite{bloch,foll,east} argued  that for a second order linear diferential equation of the form,

\[ {\cal D}(\vec{r}) \Psi(\vec{r}) = E \Psi(\vec{r}), \]

\noindent where the differential operator ${\cal D}(\vec{r})$ satisfies lattice translation symmetry : ${\cal D}(\vec{r})= {\cal D}(\vec{r}+\vec{\chi})$. There also exists a  real vector $\vec{q}$, such that the solution takes the following form:

\[ \Psi_{\vec{q}}(\vec{r},E) = \exp\{i\vec{q}\cdot\vec{r}\} u_{\vec{q}}(\vec{r},E) \]

\noindent Note that the vector $\vec{q}$ is necessarily real and labels both the quantum state and its energy. Bloch applied it to the Kohn-Sham equation for a crystalline solid, where

\[ {\cal D}(\vec{r}) = -(1/2) \nabla^2 + V(\vec{r}) \quad \mbox{with } V(\vec{r})=V(\vec{r}+\vec{\chi}) \]

\begin{figure}[t!]
 \centering
\framebox{\includegraphics[width=8.0cm,height=4.5cm]{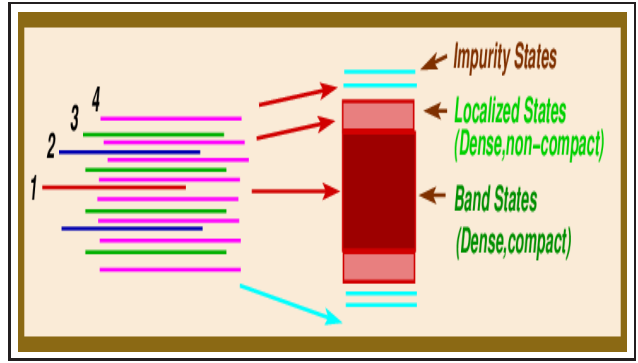}}
\caption{(Colour online) Anderson's picture of evolution of bands from discrete spectrum of finite systems.\label{Ander}}
\end{figure}

\noindent The rush to join the Bloch bandwagon was so great that many uncomfortable questions that rose from time to time with the Bloch theorem being violated were systematically swept under the formalism. From the seminal work by Anderson \cite{pwa}, Chandrasekhar \cite{nk1,nk2} and Mott \cite{nm}, it became clear that disorder which violates Bloch theorem can give rise to new phenomena where disorder cannot be regarded as a small perturbation on a crystalline background. In particular, the picture proposed by Anderson as illustrated in Fig.\ref{Ander}, is worth review. In a finite system, say an atom or a molecule, the spectrum is discrete and labelled by discrete quantum labels $\{n\}$. As the size of the molecule increases the number of discrete eigenvalues also increase.
Moreover, if $\{E_n^{(L)}\}$ is the set of eigenvalues of a molecule of size $L$, then
the eigenvalues of the system of size L exactly  those of size L-1. If, as $L\rightarrow\infty$ the spectrum of the solid is
bounded. Then by Cantor's theorem, the asymptotic spectrum ${\cal S}$ consists of two
parts : ${\cal S}_1$ in which the spectrum remains discrete. This part describes impurities, bound states and does not exist  in pristene systems. ${\cal S}_2$, which is a dense, compact set and represents the electron bands in a pristene solid.

\[ E_{n-1}^{(L-1)} \leq E_n^{(L)} \leq E_n^{(L-1)}\]

For the disordered system, there is a third type of spectrum ${\cal S}_3$ which is dense but not compact. This part represents the Anderson localized states. Thus, the problem arises in the case where the premises of the Bloch theorem breaks down and how do we proceed next to study such disordered solids.

\section{The Supercell technique.}

\begin{figure}[t!]
 \centering
\includegraphics[width=6.5cm,height=4.5cm]{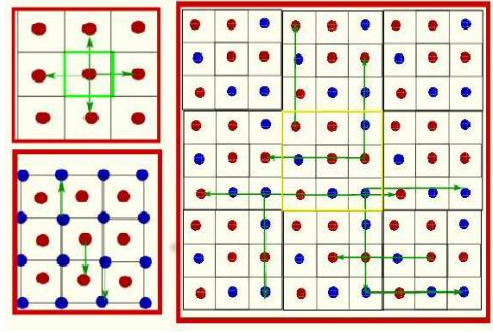}
\caption{(Colour online) (Left panel, Top) A simple primitive cell with one atom per cell. The arrangement
on the left is perfectly crystalline with lattice vectors $\{\pm a\hat{i},\pm a\hat{j}\}$. The
arrangement on the Bottom, left panel, has constructed a supercell of twice the length and breadth as the
one on the left. So there are two atoms in this unit cell. This can only be crystalline too. (Right panel) a larger super-cell  
 with lattice vectors :$\{\pm 5a\hat{i}, \pm 5a\hat{j} \}$. Now one can introduce disorder up to this scale. Beyond this the system is again periodic.\label{fig1}}
\end{figure}

A large section of physicists, ever reluctant to let go of the Bloch crutch, turned to `supercell' techniques. They expanded the crystalline unit cell to manyfold. Within this large unit cell,  the potentials are allowed to vary randomly, but periodicity is maintained between super-cells. Their justification arose from the Kohen premise of ``short-sightedness"  of electrons. In the other words, electronic structure in solids depends predominantly on   the neighborhood ${\cal R}$ within the solid.  The procedure then is to   construct the supercell around ${\cal R}$, incorporate the disorder within ${\cal R}$,  relevant in the solid and eventually sufficiently far off impose artificial lattice translation symmetry. The argument in support of this approach maintains that cutting off the infinite system at the supercell length scales and imposing periodic boundary conditions causes errors which cannot be worse than typical
finite size effects. Mathematically, it is the Resolvent or Green operator that carries the information of the spectrum. The Resolvent is defined as:
\[ {\mathbf G}(z) = (z{\bf I} - {\bf H})^{-1} \]

\noindent z is a complex number. The spectrum of {\bf H} coincides with the singularities. It has poles on the real axis and the spectrum is discrete. But for manageable sized supercells, the number of such poles is so large that they can mimic the continuous spectrum of the infinite system.  
Let us analyze the super-cell method and its drawbacks in some detail in reference to the Fig.\ref{fig1}. The top left panel shows a lattice with a primitive cell
with simply one ion-core per cell, one orbital per site in the valence sea (leading to
one band in the solid) and lattice vectors $\vec{\chi}_n = \pm a\hat{i}\ ;\ \pm  
 a\hat{j}$. The ions (and hence the potential seen by the valence electrons) have perfect lattice translation symmetry and the Bloch theorem is valid. We can
therefore introduce a real, reciprocal lattice vector $\vec{q}$ which labels both the wavefunction and the energies, which themselves are real, since the Hamiltonian is hermitian. The wave function is :

\[ \Phi_{\vec{q}}\left(\vec{r},E(\vec{q})\right) = \exp\left\{ i\vec{q}\cdot\vec{r}\right\} u_{\vec{q}}\left(\vec{r},E(\vec{q})\right) = \langle \vec{r}\vert \vec{q}\rangle \]

\noindent where  $u_{\vec{q}}(\vec{r}+\vec{\chi},E) = u_{\vec{q}}(\vec{r},E)$. 

\noindent Consider a simple, tight-binding form of the Hamiltonian: where ${\cal P}$ and ${\cal T}$ are projection and transfer operators respectively and $\vec{\chi}$ is the nearest neighbour vectors. The energy E($\vec{q}$) is
$\varepsilon + t  S(\vec{q})$ where $S(\vec{q}) = \sum_{\vec{\chi}}\  \exp(i\vec{q}\cdot\vec{\chi})$

The resolvent of {\sl H}  = (z{\sl I\ -\ H})$^{-1}$ then has a representation : 
\[ \langle\vec{q}\vert (zI-H)^{-1} \vert\vec{q}\rangle = G(\vec{q},z) = \frac{1}{z-E(\vec{q})} \]

The spectral function :
\begin{eqnarray}
 A(\vec{q},E) & = & -(1/\pi) \Im m \lim_{\eta\rightarrow 0} G(\vec{q},E+i\eta)\quad
 =  \lim_{\eta\rightarrow 0} \frac{\displaystyle\eta/\pi}{\displaystyle [E-E(\vec{q})]^2+\eta^2}  
= \delta(E-E(\vec{q})) \end{eqnarray}

  \begin{figure}[b!]
 \centering
\framebox{\includegraphics[width=8cm,height=6cm]{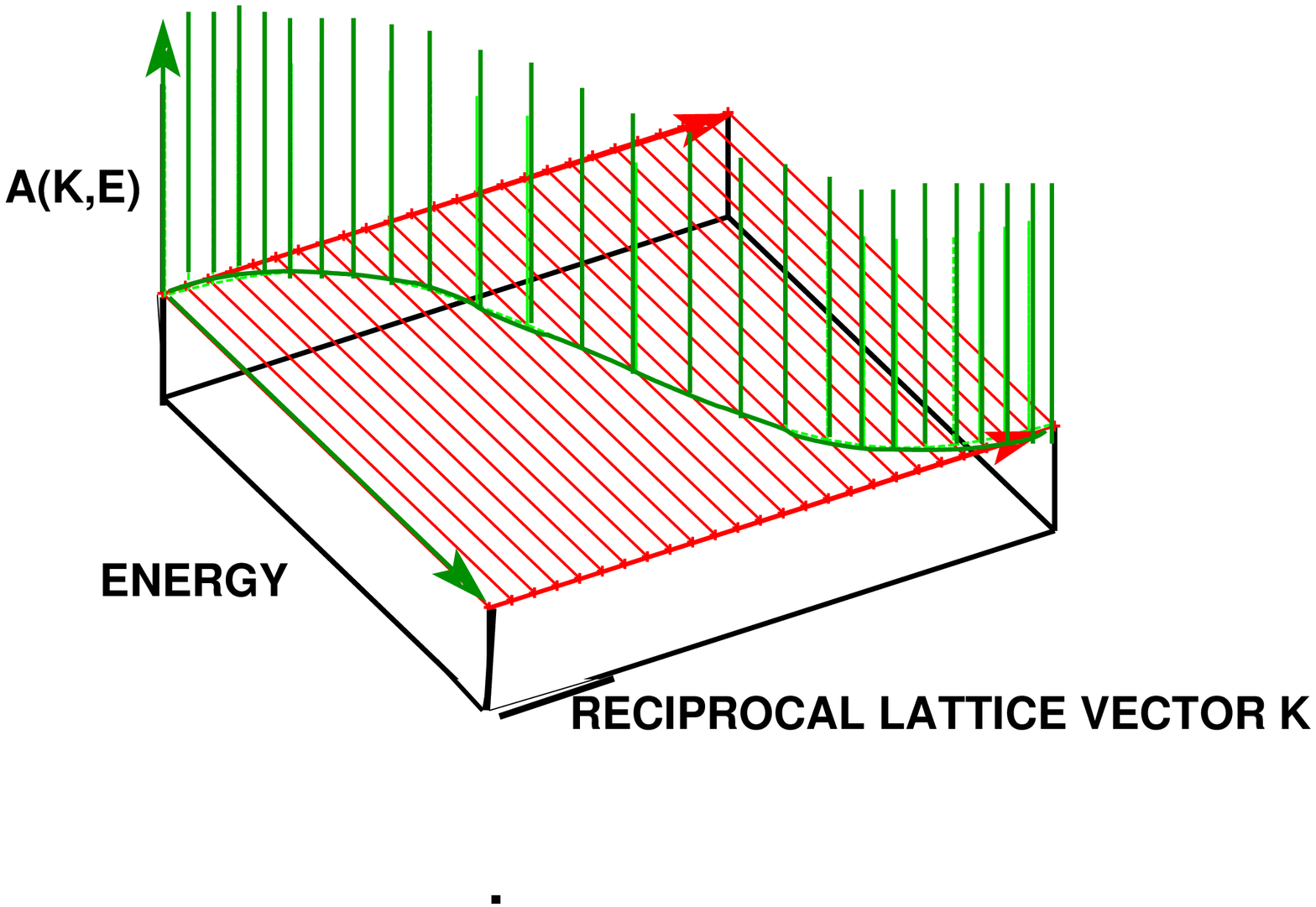}}\hskip 0.2cm
\caption{(Colour online) Spectral functions for $\vec{q}$ varying between the $\Gamma$ and $X$ points : for a perfect lattice and a primitive cell containing only one atom without disorder.\label{fig2}} 
 \end{figure}

 \begin{figure}[h!]
 \centering
\includegraphics[width=5.5cm,height=5.5cm,angle=270]{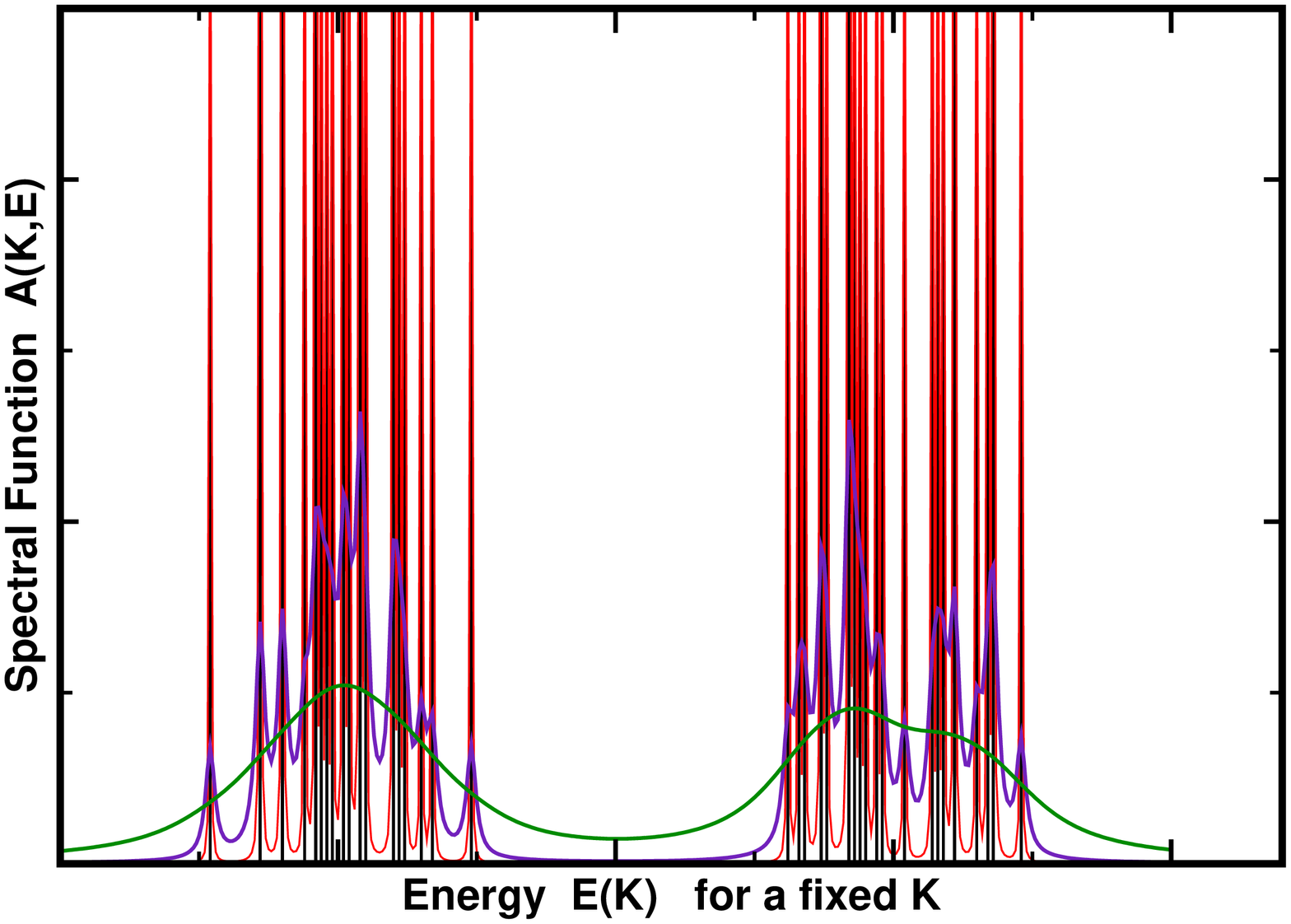}\hskip -0.5cm
\includegraphics[width=5.5cm,height=5.5cm,angle=270]{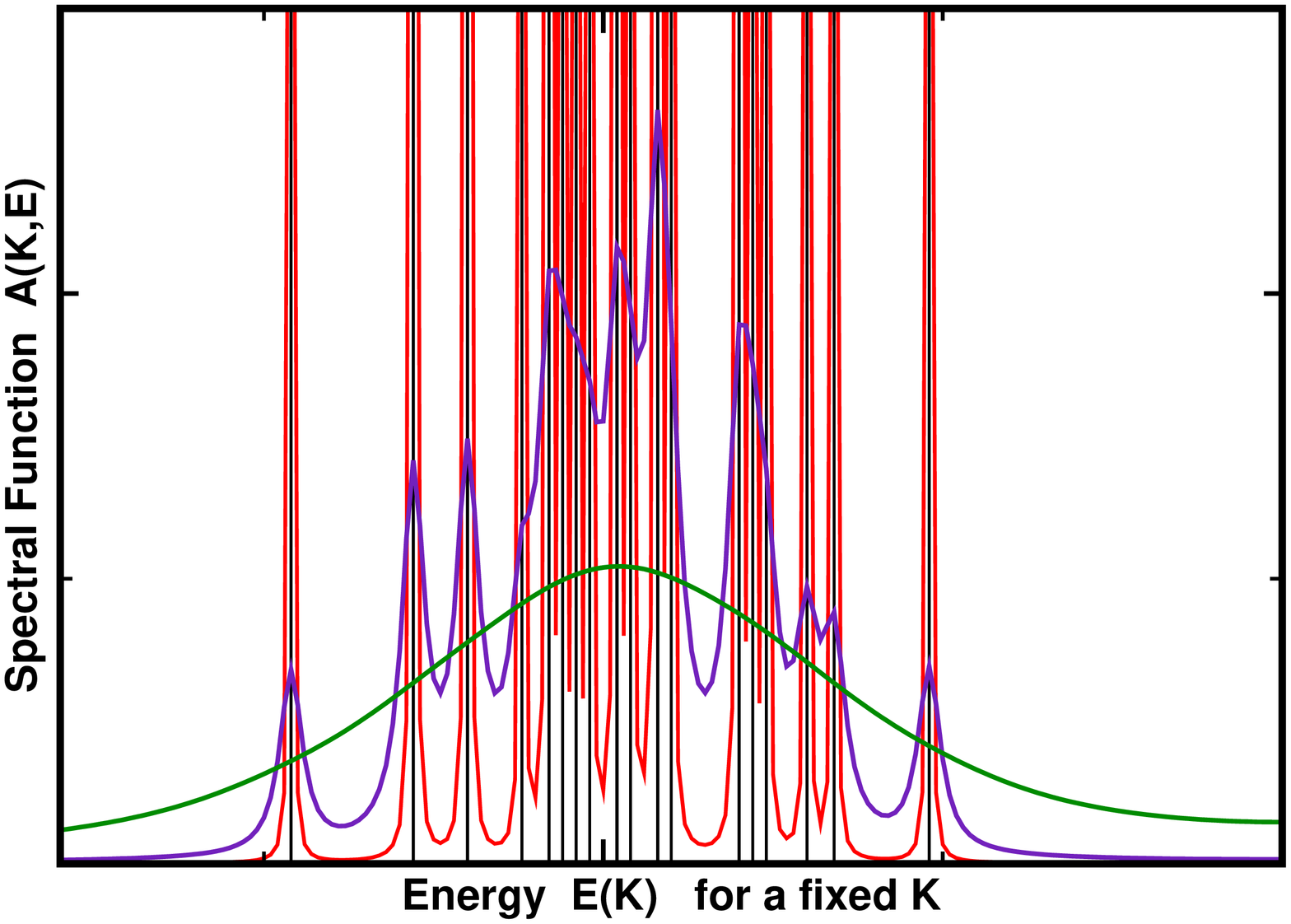} \\ \vskip -0.5cm
\includegraphics[width=5.5cm,height=5.5cm,angle=270]{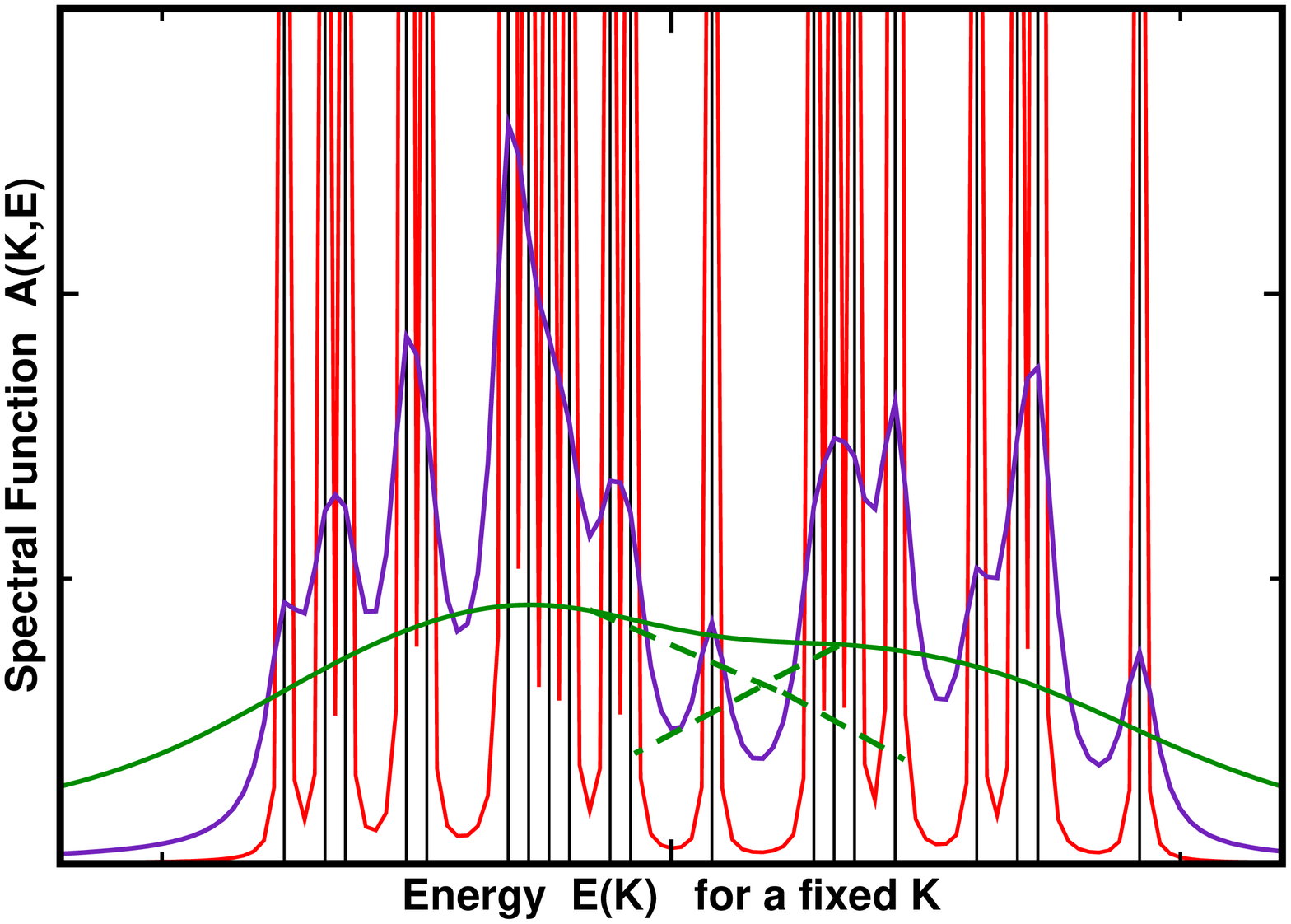}\hskip -0.5cm 
\includegraphics[width=5.5cm,height=5.5cm,angle=270]{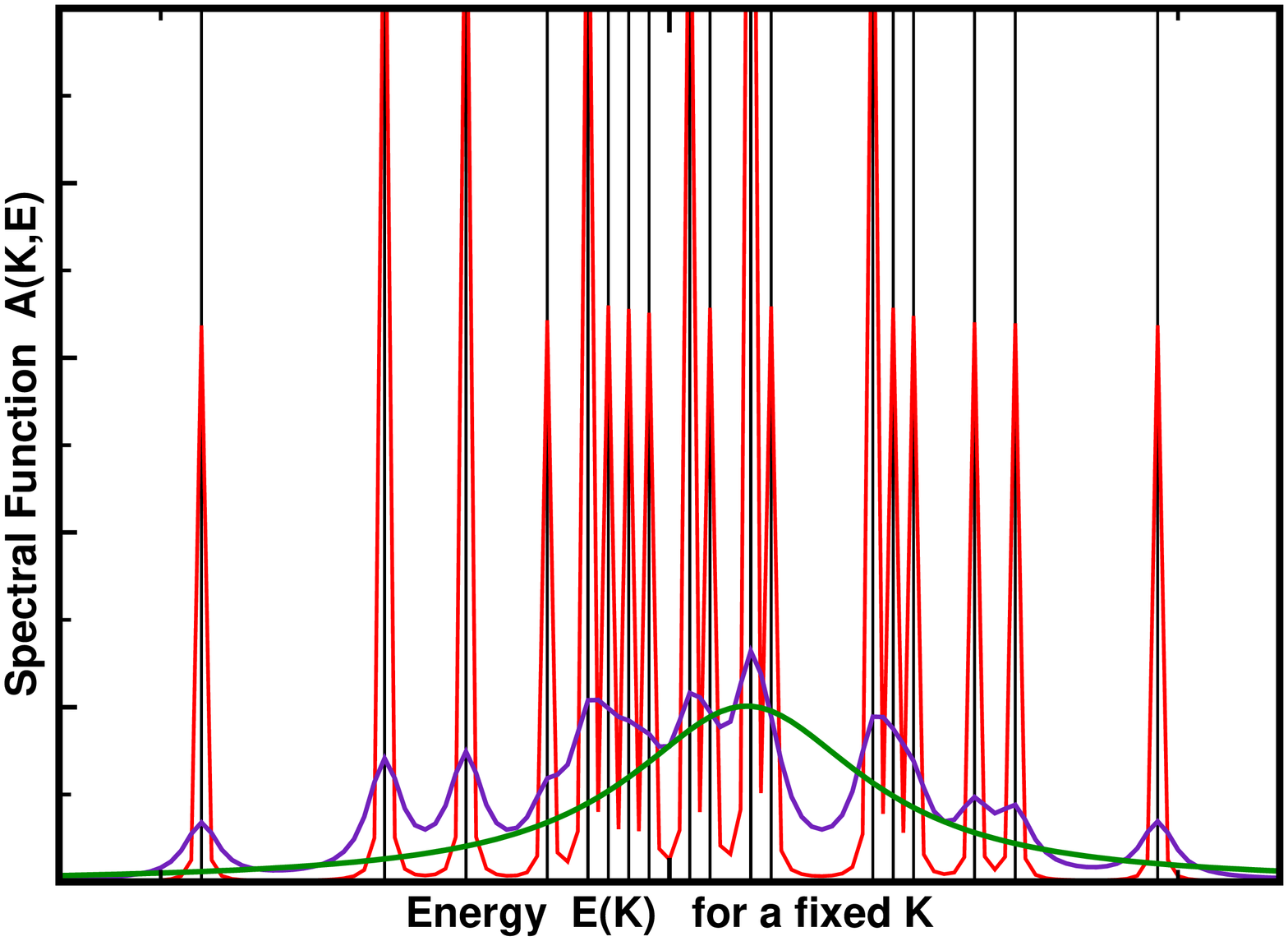} 
\caption{(Colour online) Spectral functions for different large supercells of 25 atoms. The red vertical lines mark the positions of the delta functions for a fixed $\kk$. The indigo, blue and green curves are results of various degrees of external smoothening.\label{fig3} }
\end{figure}

\begin{figure}[b!]
\centering
\includegraphics[width=4.5cm,height=3.5cm]{FIG5A.eps}
\includegraphics[width=4.0cm,height=3.5cm]{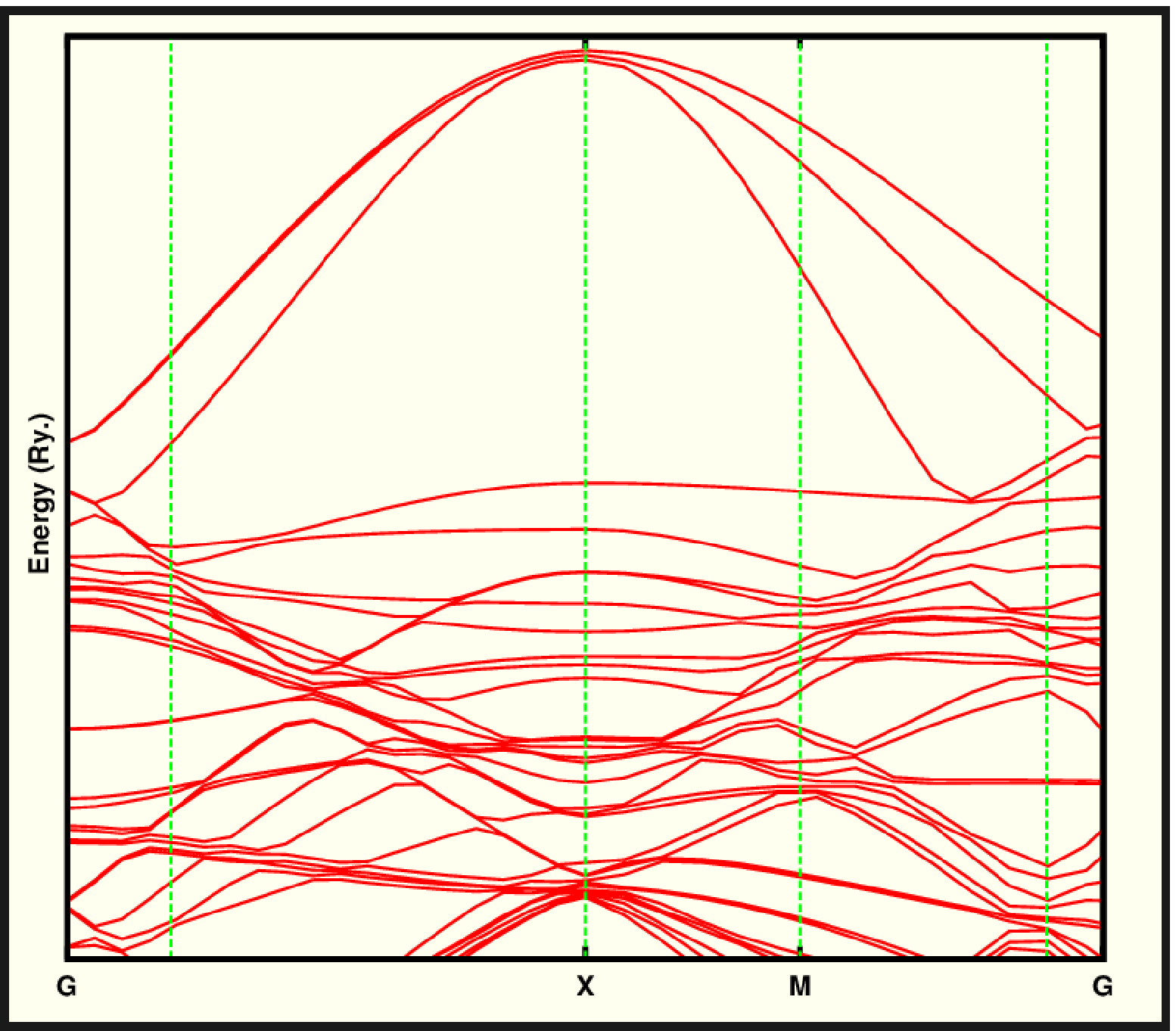}
\includegraphics[width=4.0cm,height=3.5cm]{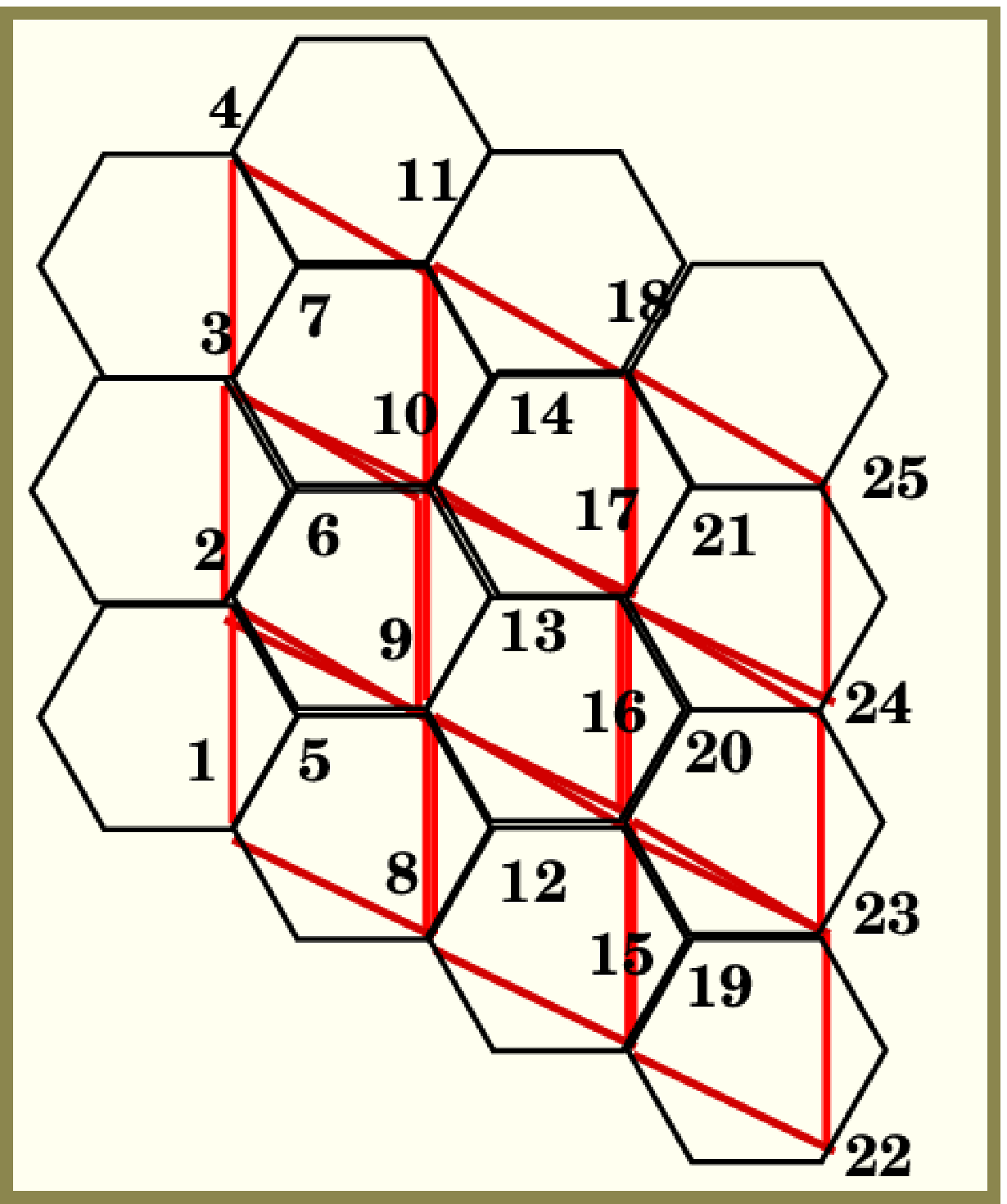}
\caption{(Left) The dispersion relation for pristine graphene. (Middle) The unsmoothened  averaged
dispersion relation  for graphene with random vacancies. This has been calculated from a super cell
made out of nine original unit cells, as shown in the panel on the right.
\label{super}}
\end{figure}

\noindent If the reciprocal vectors scale down to half of the original one, but because of lattice translation symmetry, the essentials of Bloch theorem still hold. It is noted that as long as there is lattice translation symmetry, the spectral function  always remains bunches of delta functions as shown in Fig.\ref{fig2}. The right panel of Fig.\ref{fig1} shows a super-cell with 25 lattice points and two
 	types of atoms A and B occupied the sites randomly. On the super-cell, we
 	see the random configurations, but once periodic boundary conditions are imposed, the 
 	same pattern is repeated with translation vectors $\pm 5a\hat{i}\ ;\ \pm  5a\hat{j}$. If we trace out the position of the delta functions in the energy axis, we obtain the dispersion curves for the system. For a fixed value of $\vec{q}$ this has a delta function at $E=E(\vec{q})$. A part of the spectral function between $\Gamma$ and $X$ points is shown in the left panel of Fig.\ref{fig3}. Note from Fig.\ref{fig3} and \ref{super} that the shape of the spectral density depends
upon the exact value of the externally applied imaginary part of $z$. The spectral function is still a bunch of delta functions and clustered around some average energies $\ll E_{n}(\vec{q})\gg$. The spectrum remains discrete and the spectral function remains a set of delta functions. The resolvent  $G(\vec{q},z)$ continues to have poles on the real z axis. It is customary to introduce a small complex part to $z$, which smoothens the spectral density :

\[ A(\vec{q},E-i\eta) = \frac{ \eta/\pi}{(E-E_{n}(\kk))^2+\eta^2}\]

\noindent and causes the ``bands" to gain width \cite{am2, am3}. The complex part of z is given ``by hand", like an external parameter which,
in the pure supercell method, is not obtained from first principles. The imaginary part of this complex ``self-energy" which we add to z is directly related to the inverse of a disorder induced lifetime which can be measured through neutron-scattering experiments. This is one more reason we should obtain the lifetime from first principles and not from parametrization. 
 However, if we could devise a method of mimicking the properties of the infinite sized system from those of the finite sized, but large cell and not
introduce an indeterminate parameter from outside, this approach will be
on firm ground.\\


\section{The Augmented Space Formalism}

The first successful attempts at self-consistent and first-principles estimation of the ``self-energy" culminated in the coherent potential approximation (CPA) \cite{sov,vel,am2}.  The resulting averaged resolvent maintained all the necessary herglotz\footnote{A function f(z) is herglotz if (i) f(z) has singularities only on the real z-axis. (ii) In he rest of the Argand plane f(z) is analytic with 
Sgn(Im f(z))=- Sgn(Im(z)) and (iii) f(x)$\rightarrow\pm\infty$ as x$\rightarrow \pm\infty$} analytic properties of the exact resolvent. The CPA was a single site formalism and any kind of multi-site correlated or uncorrelated randomness or extended defects could not be addressed by it. A majority of the earlier  attempts at generalizing the CPA led to a serious debate. In 1973, Mookerjee \cite{am1}
proposed a way out and introduced the Augmented Space formalism \cite{am3, am4, am5} (described in the Appendix). Suppose we have a ordered state and we measure its properties : {\bf a,b,m}$ \ldots$.  Absence of disorder means the measurement will give unique numbers {\sl a,b,m}$\ldots$ with probability. On the other hand if {\bf A,B,M$\ldots$} are some of the disordered properties, then  we cannot associate single numbers with their measurements. We have to associate a set of numbers $\{ a_1, a_2, \ldots\}$ and a probability density. This uncertainty may be from chemical sources or quantum mechanical. Taking cue from the latter, we associate an operator with the property whose eigenvalues are the measured values and whose spectral density is the probability density of measurements. Let us take an example of Hamiltonian :  
\be H\ =\ \sum_{\R} \varepsilon(\R)\ {\cal P}_{\R} + \sum_{\R}\sum_{\R+\nv} t(\nv) {\cal T}_{\R,\R+\nv} 
\ee

$H$ is an operator in the Hilbert space ${\cal H}$ spanned by the discrete basis $\{|\R\rangle\}$.
Let $n_{\R}$ be a random variable which takes the values 0 and 1 with probabilities $x$ and $y$. Then the Hamiltonian becomes, 

\begin{eqnarray}
H &=& \sum_{\vR}\ [\varepsilon^{A} n_{\vR}+\varepsilon^{B}(1-n_{\vR})]\  a^\dagger_{\vR} a_{\vR} + 
\sum_{\vR}\sum_{{\vR}^\prime}\ \left\{ \rule{0mm}{4mm} t^{AA} n_{\vR}n_{{\vR}^\prime} + t^{BB} [(1-n_{\vR})(1-n_{{\vR}^\prime})]+ \ldots\right.\nonumber\\
& & \left.\rule{0mm}{4mm}t^{AB}[n_{\vR}(1-n_{{\vR^\prime}})+n_{{\vR^\prime}}(1-n_{\vR})]\right\} a^\dagger_{\vR} a_{{\vR^\prime}}
\label{ham1}
\end{eqnarray}
Take the matrix ${ N}_{\R}$ of rank 2 with ${ N}_{\R}$ =  $\left( \begin{array}{cc}
                          y & \sqrt{xy}\\
                          \sqrt{xy} & x\end{array}\right)$.  It has eigenvalues 0 and 1 with
  spectral density $x\delta(n_{\R})+y\delta(n_{\R}-1)$. This fulfills our criterion discussed
  earlier. This is the matrix representation of an operator ${\cal N}_{\R}$ which acts on the
  `configuration' space $\Psi_{\R}$ of rank 2 spanned by
    $|\alpha_{\R}\rangle = \sqrt{x}|0_{\R}\rangle + \sqrt{y}|1_{\R}\rangle$ and $|\beta_{\R}\rangle = \sqrt{y}|0_{\R}\rangle - \sqrt{x}|1_{\R}\rangle$. Then
  
  \[ {\cal N}_{\R} = y {\cal P}^{\R}_\alpha + x {\cal P}^{\R}_\beta + \sqrt{xy} {\cal T}^{\R}_{\alpha,\beta}
  \ \in \Psi_{\R} \]                        

 The set of observables $\{n_{\R}\}$  is then associated with a set operators $\{{\cal N}_{\R}\}\ \in \Pi^\otimes \Psi_{\R} $. Combining equation \ref{ham1} with the operator, we get :

\begin{eqnarray}
\widetilde{H} & = & \sum_{\R} \ll\varepsilon\gg\ {\cal P}_{\R}\otimes {\cal I} + \sum_{\R} (y-x)(\varepsilon^A - \varepsilon^B)\ {\cal P}_{\R}\otimes {\cal P}^{\R}_\beta +\sum_{\R} \sqrt{xy}(\varepsilon^A - \varepsilon^B)\ {\cal P}_{\R}\otimes {\cal T}^{\R}_{\alpha\beta}\nonumber\\
 & & + \sum_{\vec{R}}\sum_{\vec{\chi}} \ \left[\ll t(\vec{\chi})\gg {\cal I}+(y-x)t^{(1)}(\vec{\chi})({\cal P}^{\R}_\beta+{\cal P}^{\R+\chi}_\beta)+\sqrt{xy}\ t^{(2)}(\vec{\chi})({\cal T}^{\R}_{\alpha\beta}\otimes{\cal T}^{\R+\chi}_{\alpha\beta})\right]\otimes {\cal T}_{\R,\R+\chi}\nonumber\\
\label{ham2}
\end{eqnarray}

The augmented space theorem \cite{am1}  tells us :
\be  \ll G(\kk,z)\gg = \langle \kk\otimes \emptyset| (z\widetilde{I} - \widetilde{H})^{-1} | \kk\otimes\emptyset\rangle \label{eq2} 
\ee
\n where $|\emptyset\rangle = \prod^{\otimes}\ |\alpha_{\R}\rangle$
and $|\kk\rangle = (1/N) \sum_{\R}\ \exp(i\kk\cdot\R) |\R\rangle$.
We note that if the disorder is itself homogeneous : that is the probability densities are independent of $\R$, then we can once again
introduce an augmented reciprocal space for averaged quantities. Our next step would be to obtain the averaged spectral function in equation (\ref{eq2}) using the recursion method proposed by Haydock and co-workers \cite{hhk1,hhk2}. We begin recursion with the Hamiltonian as in equation \ref{ham2} :
\[
|1\}  =  |\vec{q}\otimes\emptyset\rangle \qquad \tilde{H} |1\} \ =\  [\varepsilon+t S(\vec{q})]\ |1\} + \sqrt{xy}\Delta\varepsilon\ |\kk\otimes \{\vec{R}\}\rangle
\]
\n where $\{\R\}$ is the configuration $\{\alpha_1,\alpha_2\ldots \beta_{\R}\ldots\}$.

\begin{eqnarray}
\alpha_1 & = & \{1|\widetilde{H}|1\}/\{1|1\} = \varepsilon + t S(\vec{q}) = E(\kk)\nonumber\\
|2\} & = & \sqrt{xy}\Delta\varepsilon |\kk\otimes \{\R\}\rangle\qquad
\beta^2_2 \ =\ \{2|2\}/\{1|1\} = xy(\Delta\varepsilon)^2\nonumber\\
\tilde{H}|2\} & = & (\sqrt{xy}\Delta\varepsilon) \left\{ \varepsilon \vert\kk\otimes\{\R\}\rangle +\sum_{\nv} t\exp(-i\kk\cdot\nv)|\kk\otimes\{\R-\nv\}\rangle
+\sqrt{xy}\Delta\varepsilon |\vec{q}\otimes\emptyset\rangle\right\}\nonumber\\
\alpha_2 & = & \{2|\widetilde{H}|2\}/\{2|2\} = \varepsilon \nonumber\\
|3\} & = &  (\sqrt{xy}\Delta\varepsilon)   \sum_{\nv} t\exp(-i\kk\cdot\nv)|\kk\otimes\{\R-\nv\}\rangle\nonumber\\
\beta^2_3 & = & Zt^2 \quad\quad \mbox{where Z is the number of nearest neighbors}\nonumber\\
|N\} & = & \widetilde{H}|N-1\} - \alpha_{N-1}|N-1\} - \beta^2_{N-2}|N-2\}
\end{eqnarray}

So,\[  \{1|G(z)|1\} = G(\kk,z) = 
\frac{\displaystyle{1}}{\displaystyle z - E(\kk)-\frac{\displaystyle xy\Delta\varepsilon}{\displaystyle z - \varepsilon - \frac{\displaystyle 
Zt^2}{\ddots}}} = \frac{\displaystyle 1}{g(\kk,z)^{-1} - \Sigma(\kk,z)}
\]

\n where $g(\kk,z) = 1/(z-E(\kk))$. This leads to :
\[ G(\kk,z) = g(\kk,z)+ g(\kk,z)\Sigma(\kk,z)G(\kk,z) \]
This is Dyson's equation and we immediately recognize $\Sigma(\kk,z)$ as
the self-energy we set out to calculate.

\be \Sigma(\kk,z) = \frac{\displaystyle \beta{_2}^2}{\displaystyle z - \alpha_2-\frac{\displaystyle \beta{_3}^2}{\displaystyle z - \alpha_3 - \frac{\displaystyle \beta{_4}^2}{\phantom{\ddots }\frac{\ddots}{\displaystyle z - \alpha_N - T(z)}}}}
\ee

All practical approximations will provide us with only a finite
number of $\{\alpha_n,\beta_n\},\ n=1,2 \ldots N$. From the maximum information of the finite
number of coefficients, can we find the `terminator' $T(z)$ without entering any external parameter. The first information gleaned from  $\{\alpha_n,\beta_n\},\ n\ <\ N$ is the asymptotic behaviour of the coefficients. The averaged spectral function is 
\begin{eqnarray}
A(\vec{q},z) & = & -\frac{1}{\pi}\ \Im m \ll G(\vec{q},z) \gg \nonumber \\
& = & \frac{\Sigma_{im}(\vec{q},z)/\pi}{(g^{-1}(\vec{q},z)- \Sigma_{real}(\vec{q},z))^{2}+(\Sigma_{im}(\vec{q},z))^{2}}\nonumber\\
\end{eqnarray}
This $ \Sigma_{im}(\vec{q},z)) $ = $ 1/\tau $ is the disorder induced lifetime \cite{doni} of a $ \vec{q} $ labeled quantum state.

\section{Application to doped Graphene}

    Here we have used random vacancies as defect in graphene sheet. For a random representation of carbon with vacancies, the Hamiltonian was derived self-consistently from the DFT based tight-binding linear muffin-tin orbitals augmented space recursion (TB-LMTO-ASR) package developed by our group \cite{ASR}. In the next step, N$^{th}$ Order Muffin Tin Orbital (NMTO) method was used to construct the low-energy Hamiltonian. All the calculations are done at T=0 K.

\begin{table}[h!]
\centering
\begin{tabular}{|c|c|c|c|}\hline
 $\varepsilon$  & t$(\chi_1)$ & t$(\chi_2)$ & t$(\chi_3)$  \\ \hline 
 -0.291 eV   & -2.544 eV  & 0.1668 eV  & -0.1586 eV\\
\hline
\end{tabular}
\caption{Tight-binding parameters of Hamiltonian generated by NMTO. $\chi_n$ refer to the n-th
nearest neighbor vector on the lattice.\label{tab}}
\end{table}

\begin{figure*}[b!]
 \centering
\includegraphics[width=6cm,height=7cm,angle=270]{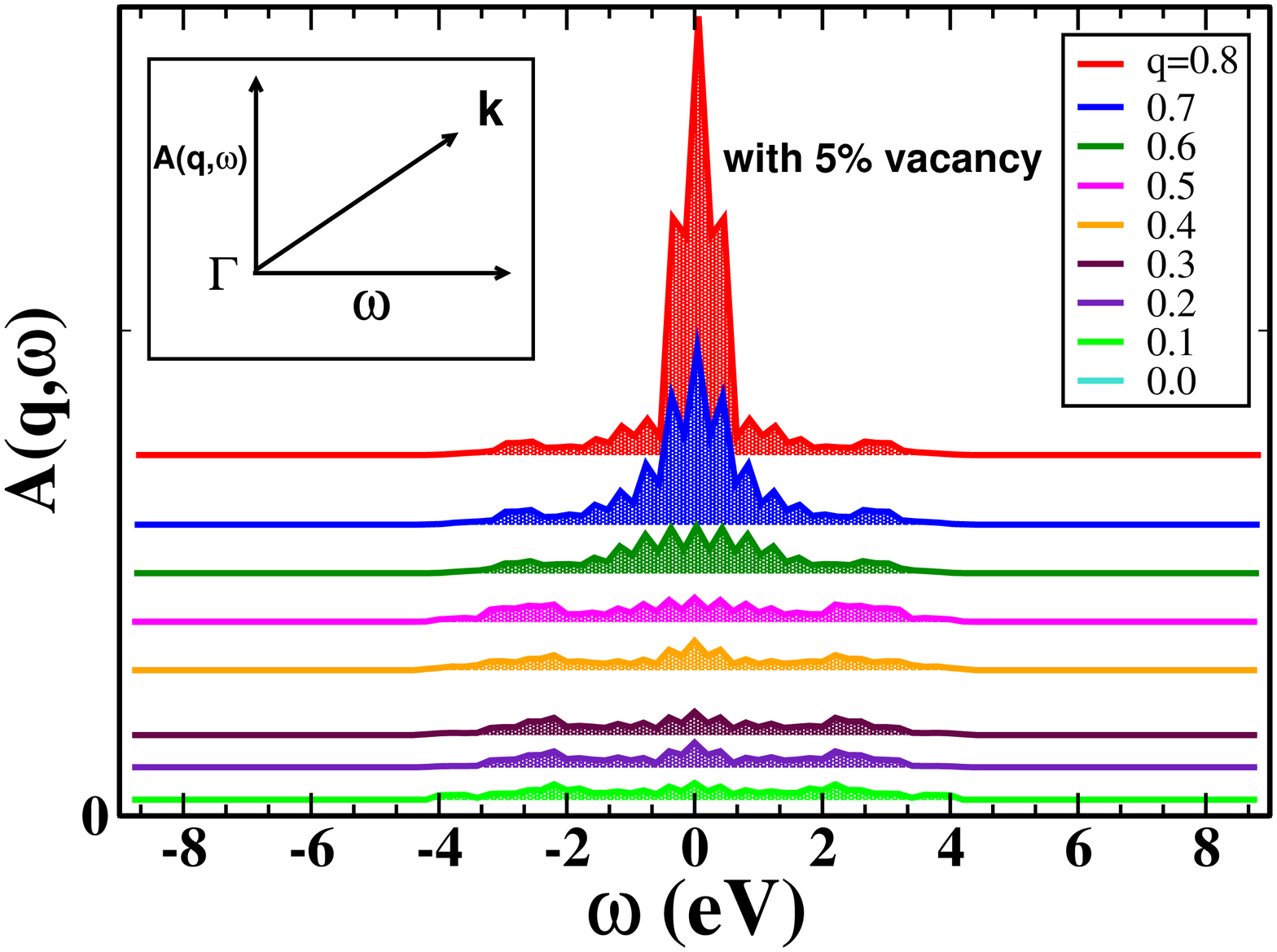}\hskip 0.5cm
\includegraphics[width=6cm,height=7cm,angle=270]{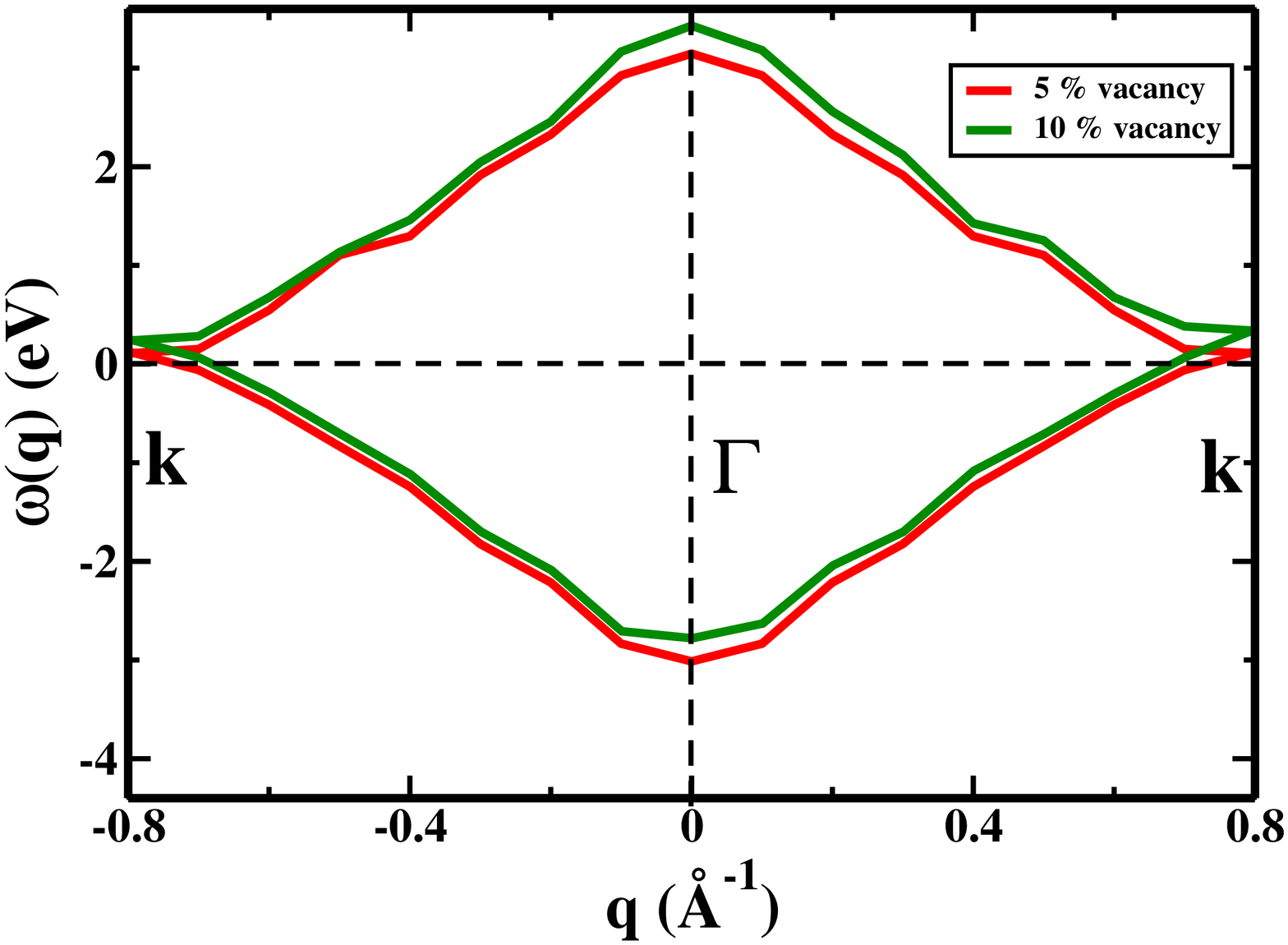}\hskip 0.2cm
\caption{(Colour online)(left) Spectral functions A($\vec{q},\omega$) for $\vec{q}$ varying between the $\Gamma$ and $K$ points  for a Graphene lattice with 5\% vacancy content. (right) The band dispersion near the Fermi level in the first Brillouin zone graphene  with random vacancies obtained by following the peak-finding process of spectral function of A($\vec{q},\omega$) described in the text. The Fermi level is at 0 eV. The linear dispersion at K point is slightly shifted upwads from Fermi level due to disorder.
\label{band}}
 \end{figure*}

			We have used augmented block recursion technique to study the energy band dispersion of doped Graphene. We have calculated the spectral function A($\kk,\omega$) as shown in Fig.\ref{band} for a selection of different wave vectors $\vec{q}$  varying from 0.0 to 0.8 along the symmetric $\Gamma-K$ with 5\% vacacy concentration. The spectral intensity of A($\kk,\omega$) is low at $\vec{q}$  = 0.0 $\AA^{-1}$ ie at $\Gamma$ point , and increases with increasing the wave vector $\vec{q}$. This is made clear by observing the peaks located at $\vec{q}$  = 0.0 ${\AA}^{-1}$ and 0.8 ${\AA}^{-1}$. The spectral intensity A($\kk,\omega$) is getting highest at dirac point ie at $K$ point. These peak  positions are shifted to lower energies with increasing wave vector $\vec{q}$  captured by our method. The peak width get broden with increasing wave vector $\vec{q}$. We obtain the band dispersions by identifying the peak positions of the spectral function A($\kk,\omega$) along the high symmetric directions in reciprocal space \cite{pk1,pk2,pk3,pk4} for doped graphene. The calculated dispersion of the $\pi$ - $\pi*$ bands is displayed in Fig.\ref{band} for doped graphene with 5\% and 10\% vacancy concentration. Doped graphene shows  linear dispersion relation above the Fermi level. The $p_z$ branches of graphene shows the typical cusp like behavior at the K point leading to its `semi-metallic' behavior. Band dispersion (BD) of disordered graphene shows linear dispersion only at K points above the Fermi level. For such a $sp^2$  bonded structure, the  $\sigma$ bond is responsible for the formation of the underlying framework. The remaining $p_z$ orbital , which is roughly vertical to the tetra or hexa-rings, binds covalently with each other and forms $\pi$ and $\pi$* bands. These $\pi$ and $\pi$* are believed to induce the linear dispersion relation near the Fermi surface. Experimentally it is observed that Silicene has a similar BS and Dirac-like fermions with regular hexagonal symmetry \cite{sili1,sili2}.

\begin{figure}[h!]
\centering
\includegraphics[width=6cm,height=7cm,angle=270]{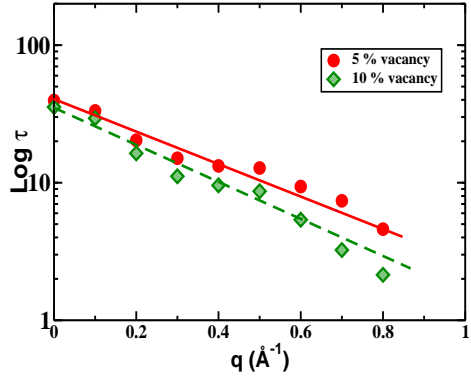}  
\caption{(Colour online) Relaxation time of graphene for $\vec{q}$ varying between the $\Gamma$ and $k$ points with random vacancies from disordered induced self energy $\Sigma_{e-e}$. \label{relax}} 
 \end{figure}

    Disordered (vacancy induced) graphene leads to two distinct momentum relaxation times ; the transport relaxation time and the quantum lifetime or the single-particle relaxation time. Our recent study does not take into account the transport relaxation time, we only calculate the single-particle relaxation time $\tau(\kk)$. From a many-body-theory viewpoint the single particle relaxation time $\tau(\kk)$  is calculated from the electron self energy of the coupled electron-impurity system. We obtain disorder induced lifetime $ \tau $ of a $ \vec{q} $ labeled quantum state from the Fourier transform of the configuration averaged self energy function. Finally, in Fig.\ref{relax} we show our calculated single particle relaxation time  $\tau(\kk)$  as a function of $\vec{q}$  (${\AA}^{-1}$) in first  Brillouin Zone for graphene. The relaxing modes or patterns are labeled by $\vec{q}$, such that the average `size' of the mode is $~O(q^{-1})$. We define the lifetime $\tau(\vec{q})$ as the time in which the disordered induced self energy amplitude drops to its $1/e$ value. The central impurity peak is increasing with increasing vacancies (disorder). We note that  Log $\tau(\kk)$ decreases almost linearly with increasing wave vector $\vec{q}$. Relaxation time $(\tau(\vec{q}))$ also decreases with increasing disorder strength. To date, most experimental studies have focused on the transport scattering time. Both the single particle relaxation time and transport scattering time are important for carrier mobilities ($ \mu $ ) of 2D graphene like systems. This two types of scattering time ratio is needed for describing scattering mechanism in such 2D systems.It is important to note that the single particle relaxation time $\tau(\kk)$ is accessible to neutron scattering experiments.

\section{Conclusion }
 In conclusion, we present a theoretical method to study the effects disorder in describing the spectral properties of graphene within the framework of Green function. We show how the topology of the Dirac dispersion and the location of Dirac point change with disorder. We also obtain disorder induced lifetime without external parameter fitting in the the complex part of energy. The single particle relaxation time  $\tau(\vec{q})$ has been obtained from the broadening of quantum states due to disorder. Disorder-induced broadening is related to the scattering length (or life-time) of Bloch electrons. We believe that such a study may provide a useful insight into the scattering mechanism of doped graphene. Finally, we believe that our framework is suitable for the study of the effects of disorder in other 2D materials such Silicene, boron nitride mono layer.

\section*{Acknowledgements}
The authors would like to thank Dhani Nafday and Tanusri Saha-Dasgupta (SNBNCBS, Kolkata) for providing us with the
Hamiltonian parameters calculated by NMTO code. We would like to thank Prof. O.K. Anderson, Max Plank Institute, Stuttgart, Germany, for his kind permission to use TB-LMTO code developed by his group.

\onecolumn
\section*{Appendix}

The augmented space method was first introduce by Mookerjee \cite{am2,am3,am1,am4} is a feasible technique for carrying out configuration averaging in disordered systems. Here we shall discuss briefly about this formalism.
Let us consider a set of independently distributed binary random variables $\{n_R\}$ with probability densities $p_{R}\{n_R\}$ and assume that the $p_{R}\{n_R\}$ has finite moment of all order. Clearly it is a reasonable assumption for almost all physical distributions. Since probability densities are positive definite functions therefore it can always be possible to express them as spectral densities of positive definite operator N$_R$ as:
\begin{equation}
 p_R(n_R)=-\frac{1}{\pi}\ \Im m \left[\langle \uparrow_R\vert(z{\bf I}-{\bf N}^R)^{-1}\vert \uparrow_R\rangle\right]
\end{equation}
\n where $z\rightarrow n_R+i\ \delta\ ; \delta\rightarrow 0$, and $\vert \uparrow_R\rangle $ is the {\sl average} state defined in such a way that the for any related quantity containing $n_R$, $\langle \uparrow_R\vert \eta \vert \uparrow_R\rangle\ $ gives average value of $\eta$
Since the resolvent of N$^R$ \ $(z{\bf I}-{\bf N}^R)^{-1}$ is Herglotz, and $p_{R}\{n_R\}$ is assumed to be such that it has finite moment of all order exists,  so one can  expand it as a continued fractional form,
\begin{equation}
p_{i}(n_i)  = -\frac{1}{\pi}\Im m\ \ \frac{1}{\displaystyle z-\alpha_0
             -\frac{\beta_{1}^{2}}{\displaystyle z-\alpha_1}}
\label{eq3.3}
\end{equation}

For a binary distribution if $n_R$ takes the value 0 and 1 with probabilities $x, \ y=(1-x)$ then $p_R(n_R)\ =\ x\delta(n_R-1) + x\delta(n_R)$  we have : $\alpha_0=x,\alpha_1=y$ and $\beta_1=\sqrt{xy}$, and a representation of N$^R$ is
\[
\left(\begin{array}{cc}
x & \sqrt{xy}  \\
 \sqrt{xy} & y
\end{array} \right)
\]

In general if $n_R$ takes $k$ different values with probability $x_k$ , then the configuration space is spanned by $k$ {\sl states} : $\vert k\rangle$ which are the eigenstates of $N_R$ with eigenvalue $k$. In that case the {\sl
average state} $\vert\emptyset_R\rangle$, which is the equivalent of $\vert\uparrow_R\rangle$ is 
 $\sum_{k}\sqrt{x_{k}}\vert k\rangle$ where $x_k$ is the probability of the variable $N_R$ to take the value $k$. The other members of the countable basis $\vert n\rangle$ may be obtained recursively from the average state through :
\begin{eqnarray}
\vert 0\rangle &=&\vert\emptyset_R\rangle\nonumber\\
\beta_{1}\vert 1\rangle &=&{\bf N}^R\ \vert 0\rangle- \alpha_{0}\vert 0\rangle\nonumber\\
\ldots && \ldots\ \ \ \ \ \ldots\ \ \ \ \ \ldots \nonumber\\
\beta_{n}\vert n\rangle &=&{\bf N}^R\ \vert n-1\rangle- \alpha_{n-1}\vert n-1\rangle-\beta_{n-1}\vert {n-2}\rangle\nonumber
\end{eqnarray}
\n In this basis, the operator ${\bf N}^R$ thus has the traditional form,
\[
\left(\begin{array}{cccccc}
a_{0} & \beta_{1} & 0 & 0 & 0 & \ldots  \\
\beta_{1} & \alpha_{1} & \beta_{2} & 0 & 0 & \ldots  \\
0 & \beta_{2} & \alpha_{2} & \beta_{3} & 0 & \ldots  \\
0 & 0 & \beta_{3} & \alpha_{3} & \beta_{4} & \ldots \\
\ldots & \ldots & \ldots & \ldots & \ldots & \ldots  \\
\end{array} \right)
\]


Let us now consider the average  of a well-behaved function $f(\diag)$ of $\diag$. By definition 

\begin{equation} 
\ll f(\diag)\gg   \eq  \int f(\diag)p(\diag) d\diag 
\eq \oint\; f(z)\; g(z)\; dz
\end{equation}

\n The integral is taken over a closed contour enclosing the singularities of $g(z)$ but not any of $f(z)$.
We assume here that $f(z)$ is well behaved, in the sense that it has no singularities in the neighbourhood
of a singularity of $g(z)$. We  now expand the function
$g(z)$ in the basis of its eigenstates $\{ \vert \mu \rangle\}$ of $N_{i}$. 
These may be either discrete or continuous. This expansion  can be written as a Stielje's 
integral in terms of the spectral density function $\rho(\mu)$ 
of  $N_{i}$

\begin{eqnarray}
 \ll f(\diag)\gg  &\eq & \int d\rho(\mu) \langle \emptyset  \vert \mu\rangle  \left[ \oint f(z) (z-\mu)^{-1}
\right] \langle \mu \vert\emptyset \rangle\nonumber\\
& \eq &  \langle \emptyset \vert \left[ \int d\rho(\mu)\; \vert\mu\rangle\; f(\mu)\;\langle\mu\vert \right]  \vert \emptyset\rangle 
\end{eqnarray}

The second line requires the function to be  well behaved at infinity. The expression in brackets on the right
side of the bottom equation is, by definition, the operator  $f(N_{R})$. It is the same functional of
 $N_{R}$ as $f(\diag)$ was of $\diag$. For example, if $f(\diag)$ is $\diag^{2}$ then  $f(N_{R})$ is
 $N^{2}_{R}$ . 
This yields the central equation of the augmented space theorem :
\begin{equation}
\ll f(\diag)\gg  \eq  \langle\emptyset\vert  f( N_{R}) \vert\emptyset\rangle \end{equation}
\n The result is significant, since we have reduced the calculation of averages to one of obtaining a particular
matrix element of an operator in the configuration space of the variable. Physically, of course, the 
augmented Hamiltonian is the {\sl collection} of all Hamiltonians. If we wish to carry out the configuration averaging of, say, the Green function element :  
\be G_{RR}(z) \eq \langle R \vert \left( zI \mns H(\{\ddiag_{R^{\prime}}\})\right)^{-1}\vert R \rangle  \ee
\n The augmented space theorem leads to :
\begin{equation}
\ll G_{RR}(z)\gg  \eq \langle R\otimes\emptyset\vert \left( z\~I \mns \widetilde{H}(\{\widetilde{N_{R^{\prime}}}\})\right)^{-1}
\vert R\otimes\emptyset\rangle
\end{equation}
where is the full configuration space.

 Mathematically, a new, countable, orthonormal basis set $\vert n \gg$ is generated in which
the augmented Hamiltonian is tridiagonal and is constructed  through a three term recurrence formula :

\begin{eqnarray*}
\vert 1\gg &  = & \sqrt{\frac{1}{N}}  \sum_{\vec{R}} \exp\{i\vec{q}\cdot\vec{R}\}  \ \vert\vec{R}\otimes\{\emptyset\}\gg\nonumber \\ 
\vert n+1\gg & = & \widetilde{\bf H}\vert n\gg - \alpha_{n}\vert n\gg - \beta^2_{n-1} \vert n-1\gg 
\end{eqnarray*}

\[
\alpha_n = \frac{\displaystyle \ll n  \vert \widetilde{\mathbf H}\vert n
\gg}{\displaystyle \ll n\vert n\gg} \quad,\quad     \beta^2_n = \frac{\displaystyle \ll n\vert n\gg}{\ll n-1\vert n-1\gg	} 
\]

This leads to :

\begin{equation}
\ll G(\vec{q},z)\gg = \frac{\displaystyle 1}{\displaystyle z-\alpha_1-\frac{\displaystyle \beta_2^2}{\displaystyle z - \alpha_2-\frac{\displaystyle \beta_3^2}{\quad\quad         \frac{\displaystyle \ddots}{\displaystyle  z-\alpha_N-\beta_{N+1}^2 T(z)}}}}
\end{equation}

Notice that while in a single configuration, there is no lattice translation symmetry in the Hamiltonian, if disorder is uniform, then
the full augmented space does. We can therefore talk about a configuration averaged spectral function :
 
\begin{eqnarray}
A({\vec{q}},z) & = & -\frac{1}{\pi}\ \Im m \ll G({\vec{q}},z) \gg \nonumber \\
\end{eqnarray}

Here $\ll...\gg$ indicates a configurational averaged quantity in case of a disordered systems.

\end{document}

%% file: macro_library.tex
\def\be{\begin{equation}}
\def\ee{\end{equation}}

\def\nc2{\left( \begin{array}{c} n \\ 2 \end{array}\right)}

\def\beq{\begin{equation}}
\def\eeq{\end{equation}}

\def\do{doubly }

\def\R{\vec{R}}

\def\n{\noindent}

  \def\ket{\vert \vert  \{ \emptyset \} \rangle}
  \def\ket2{\vert \vert \otimes \{ R \} \rangle}

\def\.#1{\mathaccent 95#1}
\def\^#1{\mathaccent 94 #1}
\def\~#1{\mathaccent "7E #1}

\def\eq{\enskip =\enskip}

\def\mns{\ -\ }

\def\diag{x_{_R}}
\def\ddiag{x}

  \def\ket{\vert \vert  \{ \emptyset \} \rangle}
  \def\ket2{\vert \vert \otimes \{ R \} \rangle}

\def\do{doubly }

  \def\ket{\vert \vert  \{ \emptyset \} \rangle}
  \def\ket2{\vert \vert \otimes \{ R \} \rangle}

  \def\ket{\vert \vert  \{ \emptyset \} \rangle}
  \def\ket2{\vert \vert \otimes \{ R \} \rangle}

\def\.#1{\mathaccent 95#1}
\def\^#1{\mathaccent 94 #1}
\def\~#1{\mathaccent "7E #1}

\def\eq{\enskip =\enskip}

%% file: Paper.bbl
\begin{thebibliography}{99}
\bibitem{bloch} Bloch, F. , Z. Physik {\bf 52} 555 (1928).
\bibitem{hk} Hohenberg, P. and Kohn, W., Phys. Rev. {\bf 136 } B864 (1964)
\bibitem{ks} Kohn, W. and Sham, L.J., Phys. Rev. {\bf 140} A1133 (1965)
\bibitem{hill} Hill, G.W., Acta Math. {\bf 8} 1 (1886).
\bibitem{floq} Floquet, G., Annales de l'École Normale Supérieure {\bf 12} 47 (1883) 
\bibitem{lyap}  Lyapunov, A.M. , in ``The General Problem of the Stability of Motion." London: Taylor and Francis.(1992) Translated by A. T. Fuller from Edouard Davaux's French translation (1907) of the original Russian dissertation (1892).
\bibitem{foll} F\'oll, H. in "Periodic Potentials and Bloch's Theorem – lectures in "Semiconductors I", The University of Kiel (1976)
 \bibitem{east} Eastham, M.S.P. , in ``The Spectral Theory of Periodic Differential Equations."  Edinburgh: Scottish Academic Press (1973).    
    
   
    \bibitem{pwa} Anderson, P.W. , Phys. Rev. {\bf 109} 1492 (1958)
    \bibitem{nk1} Kumar, N. and Jayannavar, A. , Phys. Rev. Lett. {\bf 48} 553 (1982)
    \bibitem{nk2} Tit, N., Pradhan, P. and Kumar, N., Phys. Rev. {\bf B49} 14715 (1994)
    \bibitem{nm} Mott, Sir N. in ``Conductance in Non-Crystalline Materials" (Clarendon Press) (1986)
   
    
    \bibitem{am2} Mookerjee, A., J. Phys. C : Solid State Phys {\bf 8}   29 (1975)
    \bibitem{am3} Mookerjee, A., J. Phys. C : Solid State Phys {\bf 8} 1524 (1975)
    
    \bibitem{sov} Soven, P., Phys. Rev. {\bf 156} 809 (1967)
    \bibitem{vel} Kopernik, K. , Eschrig, H., Velick\'y, B. and Hayn, R., Phys. Rev. {\bf B55}
    (1997)
\bibitem{am1} Mookerjee, A., J. Phys. C : Solid State Phys {\bf 6} 1340 (1973)
\bibitem{am4} Mookerjee, A., J. Phys. C : Solid State Phys {\bf 8} 2688 (1975)
    \bibitem{am5} Mookerjee, A., J. Phys. C : Solid State Phys {\bf 8} 2943 (1975)
\bibitem{hhk1} Haydock, R. , Heine, V. and  Kelly, M.J.,  J. Phys. C: Solid State Phys 5 (1972) 2845.
\bibitem{hhk2} Haydock, R. in  ``Solid State Physics", edited by H. Ehrenreich,F. Sietz, D. Turnbull, Academic, New York, 35 (1980).
\bibitem{doni} S. Doniach, E. H. Sondheimer \textit{Green's Functions for Solid State Physicists} Imperial College Press, (1998), chapter-V,137
\bibitem{ASR} A . Mookerjee, in A . Mookerjee, D.D Sarma (Eds.), Electronics Structure of Clustures, Surfaces and Disordered Solids, Taylors Francis , 2003
\bibitem{pk1} B. Skubic, J. Hellsvik, L. Nordström, and O. Eriksson,  \textit{J. Phys.Condens.Matter} \textbf{20}, 31 (2008)
\bibitem{pk2} X.Tao, D.P.Landau, T.C.Schulthess, and G.M.Stocks, \textit{Phys.Rev. Lett.} \textbf{95}, 087207 (2005)
\bibitem{pk3} K. Chen and D. P. Landau, \textit{Phys. Rev.B}  \textbf{49}, 3266 (1994)
\bibitem{pk4} Anders Bergman,Andrea Taroni, Lars Bergqvist, Johan Hellsvik, Björgvin Hjörvarsson, and Olle Eriksson \textit{Phys. Rev.}B \textbf{81}, 144416 (2010) DOI: 10.1103/PhysRevB.81.144416
\bibitem{sili1} S. Cahangirov, M. Topsakal, E. Aktu ̈rk, H. Sahin, and S. Ciraci, \textit{Phys. Rev. Lett.} \textbf{102}, 236804 (2009).  
\bibitem{sili2} P. De Padova, C. Quaresima, C. Ottaviani, P. M. Sheverdyaeva, P. Moras, C. Carbone, D. Topwal, B. Olivieri, A. Kara, H. Oughaddou, B. Aufray, and G. Le

    
    
\end{thebibliography}
